\documentclass[aps,pra,reprint,showpacs,superscriptaddress,notitlepage,twocolumn]{revtex4-1}%
\usepackage{amsfonts}
\usepackage{mathrsfs}
\usepackage{amsmath}
\usepackage{amssymb}
\usepackage{graphicx}
\usepackage{color}%
\usepackage{mathtools}
\usepackage{booktabs}
\usepackage[colorlinks,linkcolor=blue,citecolor=blue,hyperindex,bookmarks=false,pdfstartview=FitH]{hyperref}

\providecommand{\U}[1]{\protect\rule{.1in}{.1in}}

\newcommand{\figpanel}[2]{\hyperref[#1]{\ref*{#1}(#2)}}

\begin{document}
\title{Complex decoherence-free interactions between giant atoms}

\author{Lei Du}
\affiliation{Center for Theoretical Physics and School of Science, Hainan University, Haikou 570228, China}
\affiliation{Center for Quantum Sciences and School of Physics, Northeast Normal University, Changchun 130024, China}
\author{Lingzhen Guo}
\affiliation{Department of Applied Physics, Nanjing University of Science and Technology, Nanjing 210094, China}
\affiliation{Max Planck Institute for the Science of Light, 91058, Erlangen, Germany}
\author{Yong Li}
\affiliation{Center for Theoretical Physics and School of Science, Hainan University, Haikou 570228, China}
\affiliation{Synergetic Innovation Center for Quantum Effects and Applications, Hunan Normal University, Changsha 410081, China}

\date{\today}

\begin{abstract} 
Giant atoms provide a promising platform for engineering decoherence-free interactions which is a major task in modern quantum technologies. Here we study systematically how to implement complex decoherence-free interactions among giant atoms resorting to periodic coupling modulations and suitable arrangements of coupling points. We demonstrate that the phase of the modulation, which is tunable in experiments, can be encoded into the decoherence-free interactions, and thus enables phase-dependent dynamics when the giant atoms constitute an effective closed loop. Moreover, we consider the influence of non-Markovian retardation effect arising from large separations of the coupling points and study its dependence on the modulation parameters. 
\end{abstract}

\maketitle

\section{Introduction}\label{sec1}
Giant atoms~\cite{fiveyear} have become a powerful quantum optical paradigm, which breaks up a longstanding wisdom that atoms are usually modeled as single points based on the electric-dipole approximation. Specifically, giant atoms can be understood as quantum emitters that are coupled to a (propagating) bosonic field at multiple separate points. As the separation distances between different coupling points are comparable to the wavelength of bosonic field, giant atoms feature a peculiar self-interference effect leading to a series of unprecedented quantum optical phenomena, including frequency-dependent Lamb shift and relaxation rate~\cite{LambAFK,GLZ2017}, unconventional bound states~\cite{oscillate,WXchiral1,ZhaoWbound,VegaPRA,YuanGA,TopoCheng,2DtopoGA,oscillate2}, advanced single-photon scatterings~\cite{DLlambda,DLprr,JiaGA1,JiaGA2,YinScattering,ZhaoWScattering,CYTcp,ZhuScattering}, non-Markovian decay dynamics~\cite{nonexp,LonghiGA,DLretard,LvGA}, and chiral light-matter interactions~\cite{AFKchiral,WXchiral2,DLsyn}, to name a few. Even more strikingly, by engineering the geometrical arrangements of the coupling points, a set of giant atoms can be made fully dissipationless but featuring field-mediated coherent interactions~\cite{NoriGA,braided,FCdeco}. This phenomenon realizes the so-called decoherence-free interaction (DFI) that has potential important applications in quantum technologies, e.g., engineering large-scale quantum networks. Although DFIs can also be realized in discrete photonic lattices by tuning the atomic frequencies within the photonic band gaps~\cite{disDFI1,disDFI2,disDFI3}, this kind of interactions, however, is typically of short range and only operates within certain bandwidths since they are mediated by overlapped atom-field bound states.

It is known that electrons can acquire path-dependent phases when traveling in a magnetic field~\cite{AB1959}, while photons and phonons are immune to physical magnetic fields due to their charge neutrality. Given this fact, many efforts have been made to create synthetic magnetic fields for bosonic systems~\cite{syn1,syn2,syn3,syn4,syn5,syn6,syn7,syn8,syn9,Roushan,JinPRL,JiaST}. While most of these seminal works have concentrated on systems where the targets (e.g., atoms and resonators) are spatially close and non-Markovian retardation effects are typically ignored, very little is known about the effect of synthetic magnetic fields in large-scale quantum networks featuring field-mediated long-range interactions. Moreover, it is natural to ask if the DFIs between giant atoms, which are the result of a virtual-photon process, can be endowed with synthetic magnetism.

In this paper, we demonstrate how to realize complex DFIs between \emph{detuned} giant atoms. By modulating the atom-field couplings (or the atomic transition frequencies) properly, the phase of the modulation can be encoded into the DFI. Such a complex DFI is tunable \emph{in situ} and leads to observable phase-dependent effects when the effective Hamiltonian of the giant atoms has a closed-loop form. We find that the non-Markovian retardation effect, which is intrinsic to giant-atom systems, only introduces finite dissipation to the atoms without affecting their dynamics qualitatively. This detrimental effect can be mitigated with a smaller modulation frequency, yet an extremely slow modulation can smear the effect of the synthetic magnetic field due to the contribution of anti-rotating-wave terms. 

\section{Model and equations}\label{sec2}

\begin{figure*}[ptb]
\centering
\includegraphics[width=11 cm]{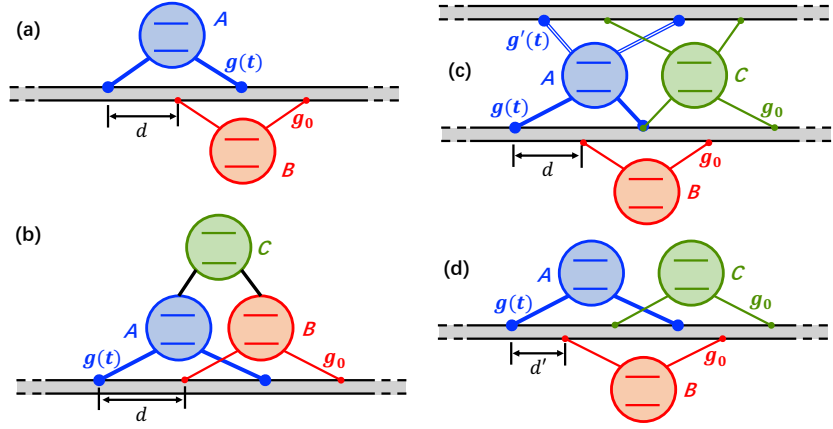}
\caption{Schematics of model architectures. (a) Two-level giant atoms $A$ and $B$ are coupled to each other via a time-dependent decoherence-free interaction. (b) A third atom $C$ is coupled directly to $A$ and $B$ to form a closed-loop atomic trimer. (c) and (d) Protected all-to-all couplings for atoms $A$, $B$, and $C$ resorting to (c) two different waveguides and (d) a single waveguide. Atoms $B$ and $C$ are assumed to be resonant with each other and detuned from atom $A$. The coupling points are equally spaced in all panels.}\label{fig1}
\end{figure*}

We start by considering a pair of two-level giant atoms (labeled as atoms $A$ and $B$, respectively), each of which is coupled to the one-dimensional waveguide at two coupling points. As shown in Fig.~\figpanel{fig1}{a}, the atom-waveguide coupling points are arranged in a braided manner that allows for a DFI between the two giant atoms~\cite{NoriGA,braided}: under certain conditions, both atoms do not dissipate into the waveguide yet there is a field-mediated coherent coupling between them. For simplicity, we assume that the coupling points are equally spaced by distance $d$ (DFIs are allowed even if the coupling points are not equally spaced). In contrast to the previous standard model where the atom-waveguide coupling strengths are constant~\cite{NoriGA,braided}, here we assume that the coupling strength $g(t)$ of atom $A$ is time dependent and the strength $g_{0}$ of atom $B$ is constant (for each atom the coupling strength is assumed to be real and identical at the two coupling points). In circuit quantum electrodynamics, such time-dependent couplings can be implemented by using a superconducting quantum interference device with tunable inductance to mediate the atom-waveguide interaction and modulate its inductance via a bias current~\cite{WXchiral2,moduscheme}. Moreover, we assume that there is a small detuning $\Delta$ between the transition frequencies of the two atoms. This detuning is crucial for realizing the synthetic magnetic field as will be shown below. With the assumptions above, the Hamiltonian of the giant-atom dimer can be written as (hereafter $\hbar=1$)
\begin{eqnarray}
H&=&H_{\text{a}}+H_{\text{w}}+H_{\text{int}}, \label{eq1}\\
H_{\text{a}}&=&\omega_{0}\sigma_{A}^{+}\sigma_{A}^{-}+(\omega_{0}+\Delta)\sigma_{B}^{+}\sigma_{B}^{-}, \label{eq2}\\
H_{\text{w}}&=&\int dk\omega_{k}a_{k}^{\dag}a_{k}, \label{eq3}\\
H_{\text{int}}&=&\int dk\left[g(t)\left(1+e^{2ikd}\right)\sigma_{A}^{+}a_{k}\nonumber\right.\\
&&\left.+g_{0}\left(e^{ikd}+e^{3ikd}\right)\sigma_{B}^{+}a_{k}+\text{H.c.}\right], \label{eq4}
\end{eqnarray}   
where $\omega_{0}$ is the transition frequency of atom $A$; $\sigma_{A}^{+}$ and $\sigma_{B}^{+}$ ($\sigma_{A}^{-}$ and $\sigma_{B}^{-}$) are the raising (lowering) operators of atoms $A$ and $B$, respectively; $\omega_{k}$ is the frequency of the waveguide field, which can be either linearly dependent on the amplitude of wave vector $k$ or linearizable around the frequency $\omega_{0}$ (with the corresponding wave vector $k_{0}$). Having in mind that the total excitation number is conserved [due to the rotating-wave approximation used in Eq.~(\ref{eq4})], the state of the model in the single-excitation subspace can be written as
\begin{equation}
\begin{split}
|\psi(t)\rangle&=\int dk c_{k}(t)a_{k}^{\dag}e^{-i\omega_{k}t}|G\rangle+\left[u_{A}(t)\sigma_{A}^{+}\right.\\
&\left.\quad\,+u_{B}(t)\sigma_{B}^{+}\right]e^{-i\omega_{0}t}|G\rangle,
\end{split}
\label{eq5}
\end{equation}
where $c_{k}$ is the probability amplitude of creating a photon with wave vector $k$ in the waveguide; $u_{A}$ and $u_{B}$ are the excitation amplitudes of atoms $A$ and $B$, respectively; $|G\rangle$ denotes that the atoms are in the ground states and there is no photon in the waveguide. Solving the Schr\"{o}dinger equation with Eqs.~(\ref{eq1})-(\ref{eq5}), one has 
\begin{eqnarray}
\dot{u}_{A}(t)&=&-i\int dk g(t)\left(1+e^{2ikd}\right)c_{k}(t)e^{-i(\omega_{k}-\omega_{0})t}, \label{eq6}\\
\dot{u}_{B}(t)&=&-i\Delta u_{B}(t)-i\int dk g_{0}\left(e^{ikd}+e^{3ikd}\right)\nonumber\\
&&\times c_{k}(t)e^{-i(\omega_{k}-\omega_{0})t}, \label{eq7}\\
\dot{c}_{k}(t)&=&-i\left[g(t)\left(1+e^{-2ikd}\right)u_{A}(t)\nonumber\right.\\
&&\left.+g_{0}\left(e^{-ikd}+e^{-3ikd}\right)u_{B}(t)\right]e^{i(\omega_{k}-\omega_{0})t}. \label{eq8}
\end{eqnarray}
By substituting the formal solution of the field amplitude (assuming that the waveguide is initially in the vacuum state)
\begin{equation}
\begin{split}
c_{k}(t)&=-i\int_{0}^{t}dt'\left[g(t')\left(1+e^{-2ikd}\right)u_{A}(t')\right.\\
&\left.\quad\,+g_{0}\left(e^{-ikd}+e^{-3ikd}\right)u_{B}(t')\right]e^{i(\omega_{k}-\omega_{0})t'}
\end{split}
\label{eq9}
\end{equation}
into Eqs.~(\ref{eq6}) and (\ref{eq7}), one can obtain the following time-delayed dynamical equations (see Appendix~\ref{appa} for more details):
\begin{eqnarray}
\dot{u}_{A}(t)&=&-\frac{2\pi g(t)}{v_{g}}\left[2g(t)u_{A}(t)+2g(t-2\tau)D_{A,2}(t)\nonumber\right.\\
&&\left.+3g_{0}D_{B,1}(t)+g_{0}D_{B,3}(t)\right], \label{eq10}\\
\dot{u}_{B}(t)&=&-i\Delta u_{B}(t)-\frac{2\pi g_{0}}{v_{g}}\left[2g_{0}u_{B}(t)+2g_{0}D_{B,2}(t)\nonumber\right.\\
&&\left.3g(t-\tau)D_{A,1}(t)+g(t-3\tau)D_{A,3}(t)\right], \label{eq11}
\end{eqnarray}   
where $D_{j,l}(t)=\text{exp}(il\phi)u_{j}(t-l\tau)\Theta(t-l\tau)$ ($j=A,\,B,\,...$ and $l=1,\,2,\,3$), with $\phi=k_{0}d$ and $\tau=d/v_{g}$ being the phase accumulation and the propagation time (time delay) of a photon traveling between adjacent coupling points, respectively; $\Theta(x)$ is the Heaviside step function. 

Equations~(\ref{eq10}) and (\ref{eq11}) describe the non-Markovian dynamics of the two giant atoms, revealing that the non-Markovian retardation effect depends on not only the coupling strength $g(t)$ \emph{at this moment} but also its values $g(t-l\tau)$ \emph{at earlier moments}. Such a feature arises from the multiple time delays among these coupling points. In Sec.~\ref{sec5}, we will also demonstrate this non-Markovian feature in a number of extended models as shown in Figs.~\figpanel{fig1}{b}-\figpanel{fig1}{d}, where an additional atom $C$ is coupled to $A$ and $B$ directly or in a decoherence-free manner via the waveguides. Before doing this, we would like to demonstrate how to implement complex DFIs in the giant-atom dimer discussed above.

\section{DFI in the Markovian regime}\label{sec3}

The multiple retardations in Eqs.~(\ref{eq10}) and (\ref{eq11}) make the dynamics of the giant-atom dimer a bit complicated. However, if $\tau$ is negligible compared to all the other characteristic time scales~\cite{footnote}, i.e., in the Markovian limit, Eqs.~(\ref{eq10}) and (\ref{eq11}) can be simplified to
\begin{eqnarray}
\dot{u}_{A}(t)&=&-\frac{4\pi g(t)^{2}}{v_{g}}\left(1+e^{2i\phi}\right)u_{A}(t)\nonumber\\
&&-\frac{2\pi g(t)g_{0}}{v_{g}}\left(3e^{i\phi}+e^{3i\phi}\right)u_{B}(t), \label{eq12}\\
\dot{u}_{B}(t)&=&-i\Delta u_{B}(t)-\frac{4\pi g_{0}^{2}}{v_{g}}\left(1+e^{2i\phi}\right)u_{B}(t)\nonumber\\
&&-\frac{2\pi g(t)g_{0}}{v_{g}}\left(3e^{i\phi}+e^{3i\phi}\right)u_{A}(t). \label{eq13}
\end{eqnarray}
Clearly, both atoms are dissipationless and their effective interaction is purely coherent when $\phi=(m+1/2)\pi$ ($m$ is an arbitrary integer). Now we consider cosine-type time-dependent couplings for atom $A$, i.e.,
\begin{equation}
g(t)=\Delta_{g}\cos{(\Omega t+\theta)}
\label{eq14}
\end{equation}
with $\Delta_{g}$, $\Omega$, and $\theta$ being the amplitude, frequency, and initial phase of the modulation, respectively. If $\Omega=\Delta\gg |2\pi\Delta_{g}g_{0}/v_{g}|$ and using the transformation $u_{B}(t)\rightarrow u_{B}(t)\text{exp}(-i\Delta t)$, Eqs.~(\ref{eq12}) and (\ref{eq13}) become 
\begin{eqnarray}
\dot{u}_{A}(t)&\simeq&-iG_{m}e^{i\theta}u_{B}(t), \label{eq15}\\
\dot{u}_{B}(t)&\simeq&-iG_{m}e^{-i\theta}u_{A}(t), \label{eq16}
\end{eqnarray}
where $\phi=(m+1/2)\pi$ has been assumed and $G_{m}=(-1)^{m}2\pi\Delta_{g}g_{0}/v_{g}$. One can see from Eqs.~(\ref{eq15}) and (\ref{eq16}) that the modulation phase $\theta$ is encoded into the DFI (with effective strength $G_{m}$), mimicking a synthetic magnetic flux for photons transferring between $A$ and $B$. Although the coupling phase $\theta$ can be gauged away for such a two-atom model (thus it has no particular interest in this case), it can significantly affect the dynamics of the system when a third atom is introduced to form a closed-loop trimer~\cite{DLretard,Roushan,arxivClerk,WXNJP}, as will be shown in Sec.~\ref{sec5}.   

Although the above analysis is only applicable in the single-excitation subspace, the decoherence-free nature of our model can also be illustrated by resorting to the theory of effective Hamiltonian~\cite{FCdeco,GAcollision,James2007}. As shown in Appendix~\ref{appb}, in the Markovian regime, the effective Hamiltonian of the giant-atom dimer can be given by
\begin{equation}
H_{\text{eff,dim}}=G_{m}e^{-i\theta}\sigma_{B}^{+}\sigma_{A}^{-}+\text{H.c.},
\label{eq17}
\end{equation}
which shows a complex DFI between atoms $A$ and $B$. Moreover, we have checked that the average interaction between the giant atoms and the waveguide field vanishes (thus the atoms are dissipationless) in this case.

Before proceeding, we briefly discuss the influence of the non-Markovian retardation effect on the result above. It is clear from Eqs.~(\ref{eq10}) and (\ref{eq11}) that the retardation effect arising from the non-negligible time delay $\tau$ may smear the DFI (such that the atoms are not perfectly dissipationless) and makes the dynamics much more complicated. To mitigate this detrimental effect, one can either consider a small enough $\tau$, or assume $\text{mod}(\Omega\tau,\pi)=0$ (a large enough $\Omega$) such that complete atomic decay can be prevented~\cite{DLprr2}.  

\section{Dynamics with effective decoherence-free interactions}\label{sec4}

\begin{figure}[ptb]
\centering
\includegraphics[width=8.6 cm]{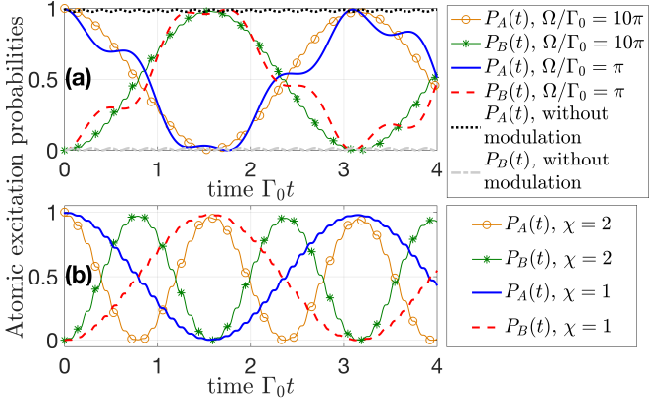}
\caption{Dynamics of atomic excitation probabilities $P_{A}(t)$ and $P_{B}(t)$ in the atomic dimer [Fig.~\figpanel{fig1}{a}] with different values of (a) $\Omega$ and (b) $\chi$. We assume $\chi=1$ in panel (a) and $\Omega/\Gamma_{0}=10\pi$ in panel (b). Moreover, we set $\Delta=\Omega$ for all lines, except for the case ``without modulation'' in panel (a): $\Omega=0$ and $\Delta/\Gamma_{0}=10\pi$ in this case. Other parameters are $\Gamma_{0}=2\pi g_{0}^{2}/v_{g}$, $\phi=\pi/2$, $\theta=0$, $\tau\Gamma_{0}=0.001$, and $|\psi(t=0)\rangle=\sigma_{A}^{+}|G\rangle$.}\label{fig2}
\end{figure} 

In this section we would like to verify the above analysis by numerically solving the time-delayed dynamical equations~(\ref{eq10}) and (\ref{eq11}) with appropriate parameters. For clarity, we use $\Gamma_{0}=2\pi g_{0}^{2}/v_{g}$ (which is the radiative decay rate of atom $B$ at each coupling point) as the unit of energies, and define $P_{A}(t)=|u_{A}(t)|^{2}$ and $P_{B}(t)=|u_{B}(t)|^{2}$ as the excitation probabilities of atoms $A$ and $B$, respectively. Moreover, we introduce a dimensionless parameter $\chi=\Delta_{g}/g_{0}$ so that the time-dependent coefficients in Eqs.~(\ref{eq10}) and (\ref{eq11}) [e.g., $2\pi g(t)g_{0}/v_{g}$] can be expressed with $\Gamma_{0}$ and $\chi$. Since we focus on the DFI of the giant atoms, hereafter we will always assume $\phi=\pi/2$ (i.e., $m=0$) and $\tau\Gamma_{0}\ll1$.

\begin{figure*}[ptb]
\centering
\includegraphics[width=13 cm]{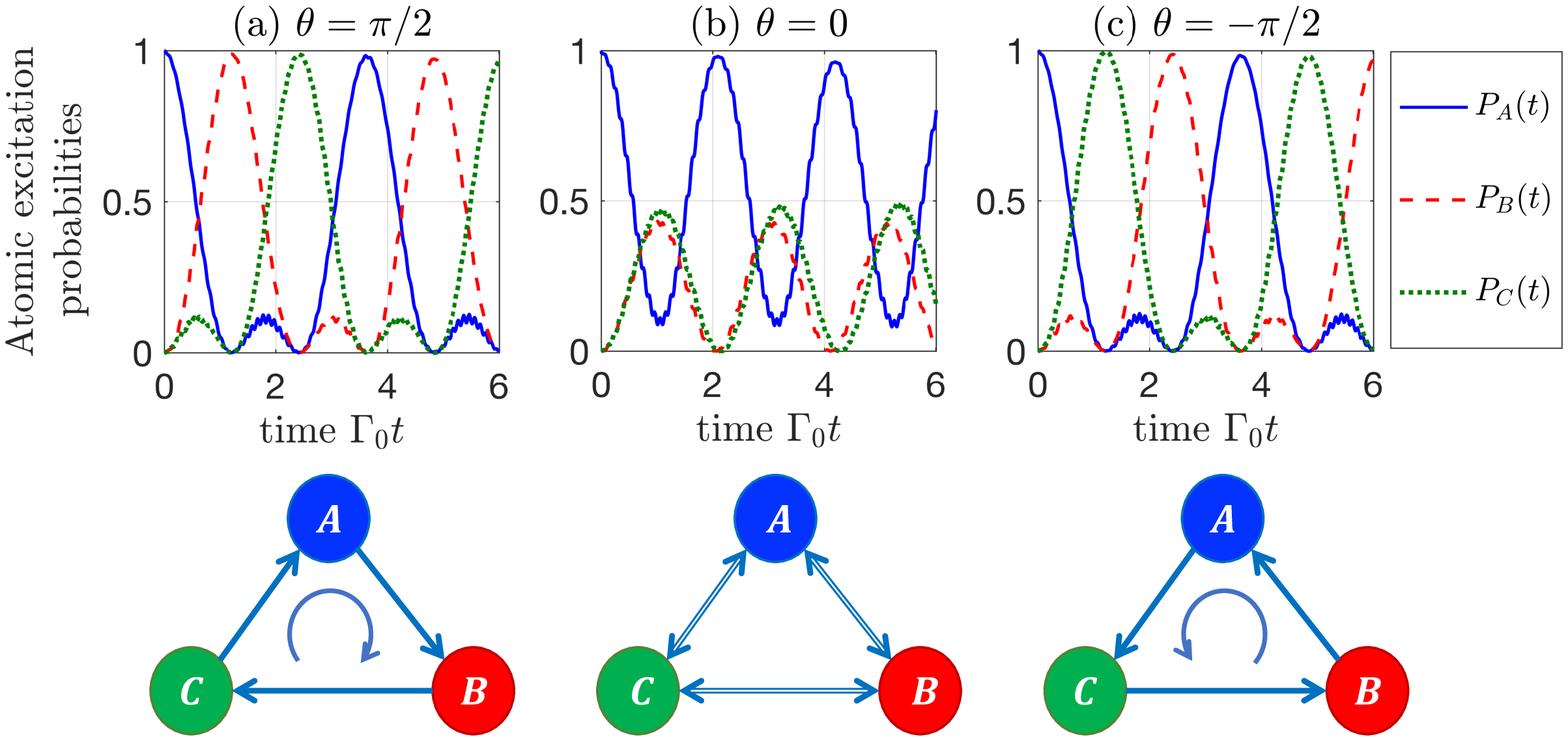}
\caption{Dynamics of atomic excitation probabilities $P_{A}(t)$, $P_{B}(t)$, and $P_{C}(t)$ in the atomic trimer [Fig.~\figpanel{fig1}{b}] with different values of $\theta$. The lower plots illustrate the excitation transfer in the trimer, corresponding to panels (a)-(c), respectively. Other parameters are $\Gamma_{0}=2\pi g_{0}^{2}/v_{g}$, $\phi=\pi/2$, $\Delta/\Gamma_{0}=\Omega/\Gamma_{0}=10\pi$, $\chi=1$, $\tau\Gamma_{0}=0.001$, and $|\psi(t=0)\rangle=\sigma_{A}^{+}|G\rangle$.}\label{fig3}
\end{figure*}

Figure~\figpanel{fig2}{a} shows the time evolutions of $P_{A}(t)$ and $P_{B}(t)$ with the initial state $|\psi(t=0)\rangle=\sigma_{A}^{+}|G\rangle$ (atom $A$ is initially excited) and with different values of modulation frequency $\Omega$. As discussed above, $\Omega=\Delta\gg |G_{m}|$ is required to justify the rotating-wave approximation [i.e., dropping high-frequency terms as in Eqs.~(\ref{eq15}) and (\ref{eq16})]. Indeed, we find that the two atoms exhibit a nearly decoherence-free excitation exchange (Rabi-like oscillation) when $\Omega$ is large enough (see, e.g., the orange line with circles and the green line with stars), while the dynamics deviate markedly from this typical form when $\Omega$ is small (see, e.g., the blue solid and red dashed lines). The Rabi-like line shapes exhibit additional tiny oscillations (thus we refer to them as ``Rabi-like'') due to the cosine-type coupling modulations. Note that the interatomic interaction almost disappears and atom $A$ exhibits a long-lived population in the absence of modulations (in this case we assume $\Omega=0$ and $\Delta/\Gamma_{0}=10\pi$ instead). This is intuitive since the two atoms have very different transition frequencies. From this point of view, the coupling modulation allows for protected interactions between \emph{detuned} giant atoms, which is significant on its own.

We also plot in Fig.~\figpanel{fig2}{b} the time evolutions of the atomic excitation probabilities with different values of $\chi$. It shows that the Rabi-like oscillation becomes faster for larger $\chi$ (i.e., larger $\Delta_{g}$), since the effective coupling strength $G_{m}$ between the two atoms is proportional to $\Delta_{g}$. This thus provides an \emph{in situ} tunable scheme for manipulating the interactions between remote quantum emitters. 

\section{Directional excitation circulation}\label{sec5} 

\begin{figure*}[pth]
\centering
\includegraphics[width=15 cm]{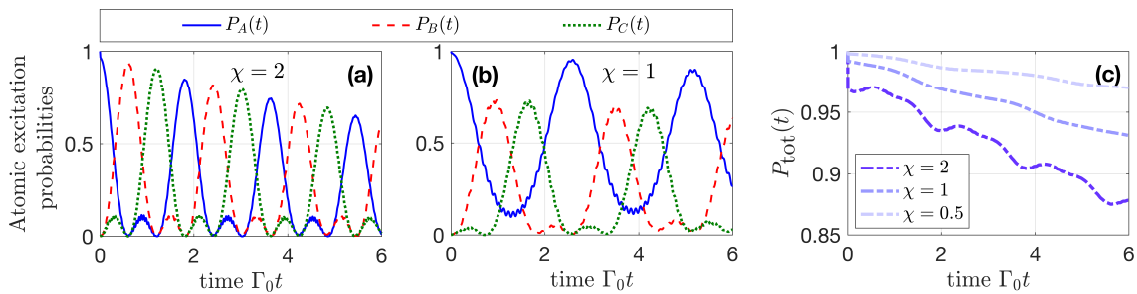}
\caption{(a, b) Dynamics of atomic excitation probabilities $P_{A}(t)$, $P_{B}(t)$, and $P_{C}(t)$ in the atomic trimer [Fig.~\figpanel{fig1}{c}] with (a) $\chi=2$ and (b) $\chi=1$. (c) Dynamics of total atomic excitation probability $P_{\text{tot}}(t)$ in the atomic trimer [Fig.~\figpanel{fig1}{c}] with different values of $\chi$. Other parameters are $\Gamma_{0}=2\pi g_{0}^{2}/v_{g}$, $\phi=\pi/2$, $\Delta/\Gamma_{0}=\Omega/\Gamma_{0}=10\pi$, $\theta=\pi/2$, $\tau\Gamma_{0}=0.001$, and $|\psi(t=0)\rangle=\sigma_{A}^{+}|G\rangle$.}\label{fig4}
\end{figure*}

As discussed in Sec.~\ref{sec3}, the effective coupling phase $\theta$ of the giant-atom dimer has no actual physical meaning since it can always be gauged away (indeed, such a coupling phase is sensitive to the choice of the initial time). In view of this, we consider an additional two-level atom (labeled as atom $C$, described by the ladder operators $\sigma_{C}^{\pm}$ and excitation amplitude $u_{C}$) coupled directly to $A$ and $B$, forming a closed-loop trimer as shown in Fig.~\figpanel{fig1}{b}. To be specific, we assume: (i) atom $C$ is resonant with atom $B$ (thus it is detuned from atom $A$ by $\Delta$); (ii) atom $C$ is coupled to atom $A$ with a time-dependent coupling strength $\lambda(t)=2G_{0}\cos{(\Omega t)}$ and to atom $B$ with a constant coupling strength $G_{0}$ ($G_{0}\coloneqq G_{m=0}=\chi\Gamma_{0}$). Considering all these assumptions, the Hamiltonian describing atom $C$ and its interaction with the other atoms can be written as 
\begin{equation}
\begin{split}
H_{\text{add}}&=(\omega_{0}+\Delta)\sigma_{C}^{+}\sigma_{C}^{-}+\left[\lambda(t)\sigma_{A}^{+}\sigma_{C}^{-}\right.\\
&\left.\quad\,+G_{0}\sigma_{B}^{+}\sigma_{C}^{-}+\text{H.c.}\right]. 
\end{split}
\label{eq18}
\end{equation}
Combined with Eqs.~(\ref{eq1})-(\ref{eq5}), the dynamical equations of the trimer can be immediately obtained as
\begin{widetext}
\begin{eqnarray}
\dot{u}_{A}(t)&=&-\frac{4\pi g^{2}(t)}{v_{g}}u_{A}(t)-\frac{4\pi g(t)g(t-2\tau)}{v_{g}}D_{A,2}(t)-\frac{2\pi g(t)g_{0}}{v_{g}}\left[3D_{B,1}(t)+D_{B,3}(t)\right]-i\lambda(t)u_{C}(t), \label{eq19}\\
\dot{u}_{B}(t)&=&-i\Delta u_{B}(t)-\frac{4\pi g_{0}^{2}}{v_{g}}\left[u_{B}(t)+D_{B,2}(t)\right]-\frac{2\pi g_{0}}{v_{g}}\left[3g(t-\tau)D_{A,1}(t)+g(t-3\tau)D_{A,3}(t)\right]-iG_{0}u_{C}(t), \label{eq20}\\
\dot{u}_{C}(t)&=&-i\Delta u_{C}(t)-i\left[\lambda(t)u_{A}(t)+G_{0}u_{B}(t)\right]. \label{eq21}
 \end {eqnarray}
 \end{widetext}
If $\phi=\pi/2$, $\Omega=\Delta$, $\tau\rightarrow0$, and $g(t)=\Delta_{g}\cos{(\Omega t+\theta)}$, Eqs.~(\ref{eq19})-(\ref{eq21}) can be simplified to
 \begin{eqnarray}
\dot{u}_{A}(t)&\simeq&-iG_{0}\left[e^{i\theta}u_{B}(t)+u_{C}(t)\right], \label{eq22}\\
\dot{u}_{B}(t)&\simeq&-iG_{0}\left[e^{-i\theta}u_{A}(t)+u_{C}(t)\right], \label{eq23}\\
\dot{u}_{C}(t)&\simeq&-iG_{0}\left[u_{A}(t)+u_{B}(t)\right]. \label{eq24}
 \end{eqnarray}
In this way, the excitation can acquire a gauge-invariant phase $\theta$ when it travels along the closed loop made of the three atoms. Such a phase simulates the synthetic magnetic flux threading the closed loop (which is typically defined as $\oint\vec{A}\cdot d\vec{r}$, with $\vec{A}$ the effective vector potential and the integral performed over the closed path~\cite{syn3,syn4,syn5}) and thus leads to phase-dependent dynamics as will be shown below.

Figure~\ref{fig3} shows the dynamics governed by Eqs.~(\ref{eq19})-(\ref{eq21}) [we define $P_{C}(t)=|u_{C}(t)|^{2}$ as the excitation probability of atom $C$], with the initial state $|\psi(t=0)\rangle=\sigma_{A}^{+}|G\rangle$ and different values of $\theta$. It shows that phase $\theta$ plays a key role in this case. In particular, as shown in Figs.~\figpanel{fig3}{a} and \figpanel{fig3}{c}, directional excitation circulation~\cite{Roushan} can be observed if $\text{mod}(\theta,\pi)=\pi/2$, with the circulation direction determined by the sign of $\theta$. According to Eqs.~(\ref{eq22})-(\ref{eq24}), the excitation transfer should be symmetric if $\text{mod}(\theta,\pi)=0$. However, as shown in Fig.~\figpanel{fig3}{b}, there is a minor difference between the time evolutions of $P_{B}(t)$ and $P_{C}(t)$, which we conclude arises from the finite retardation effect between atoms $A$ and $B$. We have checked that such a difference tends to vanish as $\tau$ decreases gradually.
 

Note that the direct interactions between $C$ and the other atoms impose some limitations on the architecture of the model. For example, atoms $A$ and $B$ have to be spatially close in order to interact directly with atom $C$. In view of this, we would like to extend the above trimer to a purely giant-atom version, where all the three atoms interact with each other via waveguide-mediated DFIs. As shown in Fig.~\figpanel{fig1}{c}, atoms $A$ and $C$ exhibit a DFI through the upper waveguide, while the DFIs between them and atom $B$ are mediated by the lower waveguide. In particular, atoms $B$ and $C$ are coupled to the waveguides with identical and constant strength $g_{0}$, whereas atom $A$ is coupled to the lower and upper waveguides with different time-dependent coupling strengths $g(t)$ and $g'(t)$, respectively (for each waveguide the two couplings of $A$ are identical). For simplicity, we still assume that the coupling points are equally spaced by distance $d$ [in the lower waveguide, atoms $A$ and $C$ share a common coupling point as shown in Fig.~\figpanel{fig1}{c}]. 

The time-delayed dynamical equations of this extended model are given in Appendix~\ref{appc} [see Eqs.~(\ref{eqc1})-(\ref{eqc3})], which, under certain conditions, show a protected all-to-all interaction (i.e., all the atoms are coupled to each other via DFIs). One may argue that the protected all-to-all interaction can also be realized by using only one waveguide as shown in Fig.~\figpanel{fig1}{d}~\cite{NoriGA}. However, we do not concentrate on this model since the global coupling phase (i.e., the total synthetic magnetic flux threading the closed loop) is always zero in this case (see Appendix~\ref{appc} for more details). Hereafter, we assume $g(t)=\Delta_{g}\cos{(\Omega t+\theta)}$ and $g'(t)=\Delta_{g}\cos{(\Omega t)}$ for the model in Fig.~\figpanel{fig1}{c} so that $\theta$ plays the role of the global coupling phase.

We plot in Figs.~\figpanel{fig4}{a} and \figpanel{fig4}{b} the time evolutions of the atomic excitation probabilities in such a giant-atom trimer with $\theta=\pi/2$. When $\chi=2$, the excitation ``hops'' directionally in sequence of $A\rightarrow B\rightarrow C\rightarrow A$, similar to that in Fig.~\figpanel{fig3}{a}, yet the damping of the total atomic excitation probability $P_{\text{tot}}(t)=P_{A}(t)+P_{B}(t)+P_{C}(t)$ is enhanced due to the stronger retardation effect in this model. From Figs.~\figpanel{fig4}{a} and \figpanel{fig4}{b} one can find that the effective coupling strength between $A$ and the other atoms (which determines the transfer efficiency and the period of the circulation) can be controlled by tuning $\chi$ (i.e., tuning the modulation amplitude $\Delta_{g}$). Moreover, as shown in Fig.~\figpanel{fig4}{c}, $P_{\text{tot}}(t)$ shows a slower damping for smaller $\chi$, since the effective decay rate of atom $A$ [described by the first two terms on the right side of Eq.~(\ref{eqc1})] decreases gradually as $\Delta_{g}$ goes to zero. This can also be seen by comparing the dynamics in Figs.~\figpanel{fig4}{a} and \figpanel{fig4}{b}.

\begin{figure}[ptb]
\centering
\includegraphics[width=8.5 cm]{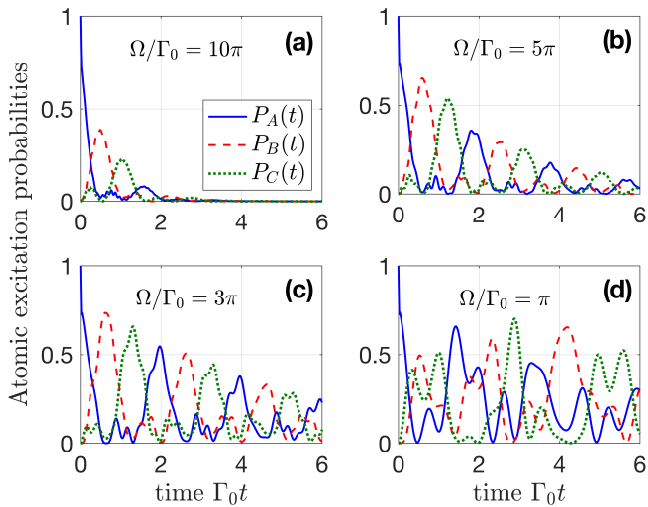}
\caption{Dynamics of atomic excitation probabilities $P_{A}(t)$, $P_{B}(t)$, and $P_{C}(t)$ in the atomic trimer [Fig.~\figpanel{fig1}{c}] with (a) $\Omega/\Gamma_{0}=10\pi$, (b) $\Omega/\Gamma_{0}=5\pi$, (c) $\Omega/\Gamma_{0}=3\pi$, and (d) $\Omega/\Gamma_{0}=\pi$. All panels in this figure share the same legend. Other parameters are $\Gamma_{0}=2\pi g_{0}^{2}/v_{g}$, $\Delta=\Omega$, $\phi=\pi/2$, $\chi=2$, $\theta=\pi/2$, $\tau\Gamma_{0}=0.01$, and $|\psi(t=0)\rangle=\sigma_{A}^{+}|G\rangle$.}\label{fig5}
\end{figure} 

Finally, we would like to demonstrate the influence of a stronger retardation effect on the present results and discuss how to mitigate this effect to some extent by tuning the modulation parameters. For relatively large $\tau$, as shown in Fig.~\figpanel{fig5}{a}, the atomic excitation probabilities become strongly damped and fall to zero rapidly, although the directional excitation circulation can still be observed. Such a rapid damping, however, can be weakened by using a smaller modulation frequency as shown in Figs.~\figpanel{fig5}{b} and \figpanel{fig5}{c} ($\Delta=\Omega$ is always satisfied). This phenomenon can be understood again from the effective decay rate of atom $A$: as shown in Eq.~(\ref{eqc1}), atom $A$ can be finally dissipationless if $g(t)=g(t-2\tau)$ and $g'(t)=g'(t-2\tau)$ [i.e., $\text{mod}(\Omega\tau,\pi)=0$], while its effective decay increases with $\Omega$ if $0<\text{mod}(\Omega\tau,\pi)\ll\pi$. However, decreasing the value of $\Omega$ also smears the directional excitation circulation since the anti-rotating-wave terms (i.e., the high-frequency oscillating terms in the effective Hamiltonian and the dynamical equations) come into play eventually. As shown in Fig.~\figpanel{fig5}{d}, the excitation transfer becomes ruleless when $\Omega$ is small enough. In other words, there is a tradeoff between the retardation-induced dissipation and the effect of synthetic magnetic field in this case.

\section{Frequency-modulation schemes}\label{sec6}  

In principle, the synthetic magnetic field can also be created by modulating the transition frequencies of the giant atoms. For example, recalling the giant-atom dimer in Fig.~\figpanel{fig1}{a}, one can assume constant and uniform coupling strengths for both atoms and a time-dependent transition frequency for atom $B$ (thus the detuning between $A$ and $B$ is time dependent). Then a complex DFI between the two atoms can be realized under certain conditions, as shown in Appendix~\ref{appd}. However, the coupling-modulation scheme shows two major advantages over the frequency-modulation one~\cite{arxivClerk}: (i) the requirements for the rotating-wave approximation to be valid are less severe in the coupling-modulation scheme; (ii) for the frequency-modulation scheme, there are many sidebands that cannot be neglected in many cases (especially when multiple frequency modulations are considered or a relatively faster modulation is employed), which may smear the DFI. Therefore we concentrate on the coupling-modulation scheme in this paper.

\section{Conclusions and outlooks}
In summary, we have demonstrated how to create a synthetic magnetic field for the effective decoherence-free Hamiltonian of giant atoms resorting to periodic coupling modulations and suitable arrangements of atom-waveguide coupling points. With our scheme one can not only realize DFIs between detuned giant atoms, but also observe phase-dependent dynamics in closed-loop chains of giant atoms. Moreover, we have considered the non-Markovian retardation effect and studied its influence on the atomic dynamics. The retardation effect does not alter the phase dependence of the dynamics qualitatively, and its resulting dissipation can be controlled via the modulation parameters within a certain range.   

The results in this paper can be applied to many applications and further investigations. For example, our scheme highlights a way towards quantum simulations of many-body systems that are subject to various gauge fields and towards engineering more high-fidelity quantum gates~\cite{NoriGA,braided}. It is also possible to generate fractional quantum Hall states of light by simply increasing the size of our models (e.g., implementing two-dimensional square or quasi-one-dimensional ladder lattices of giant atoms with tailored couplings)~\cite{Roushan}. Although in this paper we have concentrated on models made up of superconducting qubits and  microwave transmission lines, our proposal is general and can be immediately extended to other possible setups, such as quantum emitters coupled to real-space or synthetic discrete lattices. Moreover, the synthetic gauge field offers the opportunity of implementing richer topological phases based on the effective spin Hamiltonians of giant atoms~\cite{WXchiral1}.        

\section*{Acknowledgments}
We would like to thank F.~Ciccarello, Y.~Zhang and Y.-T. Chen for helpful discussions. This work was supported by the National Natural Science Foundation of China (under Grants No. 12274107 and No. 12074030).

\appendix
\begin{widetext}
\section{Time-delayed dynamical equations of the giant-atom dimer}\label{appa}

In this Appendix we demonstrate in detail how to derive the time-delayed dynamical equations (\ref{eq10}) and (\ref{eq11}) of the giant-atom dimer. By substituting Eq.~(\ref{eq9}) into Eqs.~(\ref{eq6}) and (\ref{eq7}), we have
\begin{eqnarray}
\dot{u}_{A}(t)&=&-\int_{0}^{t}dt'\int_{-\infty}^{+\infty}dk e^{-i(\omega_{k}-\omega_{0})(t-t')}\left\{2g(t)g(t')[1+\cos{(2kd)}]u_{A}(t')\nonumber\right.\\
&&\left.+g(t)g_{0}\left(e^{ikd}+2e^{-ikd}+e^{-3ikd}\right)u_{B}(t')\right\}, \label{eqa1}\\
\dot{u}_{B}(t)&=&-i\Delta u_{B}(t)-\int_{0}^{t}dt'\int_{-\infty}^{+\infty}dk e^{-i(\omega_{k}-\omega_{0})(t-t')}\left\{2g_{0}^{2}[1+\cos{(2kd)}]u_{B}(t')\nonumber\right.\\&&\left.+g(t')g_{0}\left(2e^{ikd}+e^{-ikd}+e^{3ikd}\right)u_{A}(t')\right\}. \label{eqa2}
\end{eqnarray}
If we change the integration variable as $\int_{-\infty}^{+\infty}dkf(k)\rightarrow\int_{0}^{+\infty}d\omega_{k}[f(k)+f(-k)]/v_{g}$ and write the dispersion relation of the waveguide as $\omega_{k}=\omega_{0}+\nu_{k}=\omega_{0}+(k-k_{0})v_{g}$~\cite{JTShen2005,JTShen2009}, with $k_{0}$ the wave vector corresponding to frequency $\omega_{0}$ and $v_{g}$ the group velocity of the emitted photon, Eqs.~(\ref{eqa1}) and (\ref{eqa2}) become
\begin{eqnarray}
\dot{u}_{A}(t)&=&-\frac{1}{v_{g}}\int_{0}^{t}dt'\int_{-\infty}^{+\infty}d\nu_{k}e^{-i\nu_{k}(t-t')}\left\{4g(t)g(t')[1+\cos{(2kd)}]u_{A}(t')\nonumber\right.\\
&&\left.+g(t)g_{0}\left(3e^{ikd}+3e^{-ikd}+e^{3ikd}+e^{-3ikd}\right)u_{B}(t')\right\}, \nonumber\\
&=&-\frac{2\pi}{v_{g}}\int_{0}^{t}dt'\left\{2g(t)g(t')\left[2\delta(t-t')+e^{2i\phi}\delta(t-t'-2\tau)\right]u_{A}(t')\nonumber\right.\\
&&\left.+g(t)g_{0}\left[3e^{i\phi}\delta(t-t'-\tau)+e^{3i\phi}\delta(t-t'-3\tau)\right]u_{B}(t')\right\}, \label{eqa3}\\
\dot{u}_{B}(t)&=&-i\Delta u_{B}(t)-\frac{1}{v_{g}}\int_{0}^{t}dt'\int_{-\infty}^{+\infty}d\nu_{k}e^{-i\nu_{k}(t-t')}\left\{4g_{0}^{2}[1+\cos{(2kd)}]u_{B}(t')\nonumber\right.\\
&&\left.+g(t')g_{0}\left(3e^{ikd}+3e^{-ikd}+e^{3ikd}+e^{-3ikd}\right)u_{A}(t')\right\}\nonumber\\
&=&-i\Delta u_{B}(t)-\frac{2\pi}{v_{g}}\int_{0}^{t}dt'\left\{2g_{0}^{2}\left[2\delta(t-t')+e^{2i\phi}\delta(t-t'-2\tau)\right]u_{B}(t')\nonumber\right.\\
&&\left.+g(t')g_{0}\left[3e^{i\phi}\delta(t-t'-\tau)+e^{3i\phi}\delta(t-t'-3\tau)\right]u_{A}(t')\right\}, \label{eqa4}
\end{eqnarray}
where $\phi=k_{0}d$ and $\tau=d/v_{g}$. In the last steps of Eqs.~(\ref{eqa3}) and (\ref{eqa4}), we have omitted the time-advanced terms containing $\delta(t-t'+l\tau)$ ($l=1,2,3$) since they do not contribute to the integral $\int_{0}^{t}(\cdots)dt'$. Finally, one can obtain the time-delayed dynamical equations~(\ref{eq10}) and (\ref{eq11}) by using the sifting property $\int dtf(t)\delta(t-t')=f(t')$ of $\delta$ functions.

\section{Effective Hamiltonian}\label{appb}

In this Appendix we would like to demonstrate the decoherence-free mechanism of the giant-atom dimer in Fig.~\figpanel{fig1}{a} by deriving its effective Hamiltonian. We first consider a more general situation where a set of two-level giant atoms are coupled to a common waveguide with arbitrary arrangements of coupling points. Similar to the models studied in this paper, one of the atoms (with transition frequency $\omega_{0}$; labeled as atom $A$) is detuned from the others by $\Delta$ and is coupled to the waveguide with time-dependent coupling strength $g(t)$ at two coupling points, while the other giant atoms have the same transition frequency ($\omega_{0}+\Delta$) and are coupled to the waveguide with constant coupling strength $g_{0}$. In this case, the Hamiltonian in the interaction picture can be written as
\begin{equation}
\begin{split}
V(t)&=\int dk\Big[g(t)\left(e^{-ikx_{A1}}+e^{-ikx_{A2}}\right)\sigma_{A}^{-}a_{k}^{\dag}e^{i\Delta_{k}t}+g_{0}\sum_{j,l}e^{-ikx_{jl}}\sigma_{j}^{-}a_{k}^{\dag}e^{i(\Delta_{k}-\Delta)t}+\text{H.c.}\Big]\\
&=\int dk\Big\{g(t)\left[e^{-i\varphi_{A1}}e^{i\Delta_{k}(t-\tau_{A1})}+e^{-i\varphi_{A2}}e^{i\Delta_{k}(t-\tau_{A2})}\right]\sigma_{A}^{-}a_{k}^{\dag}\\
&\quad\,+g_{0}\sum_{j,l}e^{-i\varphi_{jl}}e^{i\Delta_{k}(t-\tau_{jl})}e^{-i\Delta t}\sigma_{j}^{-}a_{k}^{\dag}+\text{H.c.}\Big\},
\end{split}
\label{eqb1}
\end{equation}
where $\Delta_{k}=\omega_{k}-\omega_{0}$. $x_{jl}$ is the position of the $l$th coupling point of atom $j$, with which we define $\tau_{jl}=x_{jl}/v_{g}$ and $\varphi_{jl}=k_{0}x_{jl}$. In the second step of Eq.~(\ref{eqb1}) we have used the linearized dispersion relation $\omega_{k}=\omega_{0}+(k-k_{0})v_{g}$. If we consider a discrete time axis $t_{n}=nT$ with the time interval $T$ short enough compared with the characteristic time of interaction, the average interaction can be defined as~\cite{FCdeco,GAcollision} 
\begin{equation}
\bar{V}=\frac{1}{T}\int_{t_{n-1}}^{t_{n}}ds V(s),
\label{eqb2}
\end{equation}
and the effective Hamiltonian of the giant atoms can be given by
\begin{equation}
H_{\text{eff}}=\frac{-i}{2T}\int_{t_{n-1}}^{t_{n}}ds\int_{t_{n-1}}^{s}ds' [V(s),V(s')].
\label{eqb3}
\end{equation}
To realize decoherence-free Hamiltonians, it is necessary to fulfill the condition $\bar{V}=0$. Now if we consider the giant-atom dimer in Fig.~\figpanel{fig1}{a} with cosine-type time-dependent couplings $g(t)=\Delta_{g}\cos{(\Delta t+\theta)}$ for atom $A$ and perform the transformation $a_{k}\rightarrow a_{k}\text{exp}(-i\Delta t)$, Eq.~(\ref{eqb1}) becomes
\begin{equation}
\begin{split}
V(t)&=\int dk\Big\{\frac{\Delta_{g}}{2}\left[e^{i\Delta_{k}t}+e^{-2i\phi}e^{i\Delta_{k}(t-2\tau)}\right]\left(e^{-i\theta}+e^{2i\Delta t}e^{i\theta}\right)\sigma_{A}^{-}a_{k}^{\dag}\\
&\quad\,+g_{0}\left[e^{-i\phi}e^{i\Delta_{k}(t-\tau)}+e^{-3i\phi}e^{i\Delta_{k}(t-3\tau)}\right]\sigma_{B}^{-}a_{k}^{\dag}+\text{H.c.}\Big\},
\end{split}
\label{eqb4}
\end{equation}
where we have assumed $\{x_{A1},\,x_{B1},\,x_{A2},\,x_{B2}\}=\{0,\,d,\,2d,\,3d\}$, $\phi=k_{0}d$, and $\tau=d/v_{g}$ as defined in the main text. Substituting Eq.~(\ref{eqb4}) into Eq.~(\ref{eqb3}) we can obtain the effective Hamiltonian of the giant-atom dimer, i.e.,
\begin{equation}
\begin{split}
H_{\text{eff,dim}}&\simeq\frac{-i}{2T}\frac{2\pi\Delta_{g}g_{0}}{v_{g}}\int_{t_{n-1}}^{t_{n}}ds\int_{t_{n-1}}^{s}ds'\Big\{\Big[2e^{i\phi}[\delta(s'-s+\tau)-\delta(s-s'+\tau)]\\
&\quad\,+e^{3i\phi}[\delta(s'-s+3\tau)-\delta(s-s'+3\tau)]+e^{-i\phi}[\delta(s'-s-\tau)-\delta(s-s'-\tau)]\Big]e^{-i\theta}\sigma_{B}^{+}\sigma_{A}^{-}+\text{H.c.}\Big\}\\
&=\frac{-i\pi\Delta_{g}g_{0}}{v_{g}}\left[(2e^{i\phi}+e^{3i\phi}-e^{-i\phi})e^{-i\theta}\sigma_{B}^{+}\sigma_{A}^{-}+\text{H.c.}\right],
\end{split}
\label{eqb5}
\end{equation}
where we have assumed that all the time delays $l\tau$ are negligible compared to $T$ (Markovian regime) and have dropped the high-frequency oscillating terms containing $\text{exp}(2i\Delta t)$. When $\phi=(m+1/2)\pi$, the effective Hamiltonian becomes
\begin{equation}
H_{\text{eff,dim}}=G_{m}e^{-i\theta}\sigma_{B}^{+}\sigma_{A}^{-}+\text{H.c.}
\label{eqb6}
\end{equation}
with $G_{m}=(-1)^{m}2\pi\Delta_{g}g_{0}/v_{g}$, which shows a complex DFI between atoms $A$ and $B$. Moreover, one can see from Eqs.~(\ref{eqb2}) and (\ref{eqb4}) that the average interaction between the giant atoms and the waveguide field vanishes (i.e., $\bar{V}=0$) in this case. 

\section{Dynamical equations of the models in Figs.~\figpanel{fig1}{c} and \figpanel{fig1}{d}}\label{appc}

For the giant-atom trimer in Fig.~\figpanel{fig1}{c}, the time-delayed dynamical equations of the atomic excitation amplitudes can be immediately given by 
\begin{eqnarray}
 \dot{u}_{A}(t)&=&-\frac{4\pi [g^{2}(t)+g'^{2}(t)]}{v_{g}}u_{A}(t)-\frac{4\pi[g(t)g(t-2\tau)+g'(t)g'(t-2\tau)]}{v_{g}}D_{A,2}(t)-\frac{2\pi g(t)g_{0}}{v_{g}}\left[3D_{B,1}(t)+D_{B,3}(t)\right]\nonumber\\
 &&-\frac{2\pi g(t)g_{0}}{v_{g}}\left[u_{C}(t)+2D_{C,2}(t)+D_{C,4}(t)\right]-\frac{2\pi g'(t)g_{0}}{v_{g}}\left[3D_{C,1}(t)+D_{C,3}(t)\right], \label{eqc1}\\
\dot{u}_{B}(t)&=&-i\Delta u_{B}(t)-\frac{4\pi g_{0}^{2}}{v_{g}}\left[u_{B}(t)+D_{B,2}(t)\right]-\frac{6\pi g(t-\tau)g_{0}}{v_{g}}D_{A,1}(t)\nonumber\\
&&-\frac{2\pi g(t-3\tau)g_{0}}{v_{g}}D_{A,3}(t)-\frac{2\pi g_{0}^{2}}{v_{g}}\left[3D_{C,1}(t)+D_{C,3}(t)\right], \label{eqc2}\\
\dot{u}_{C}(t)&=&-i\Delta u_{C}(t)-\frac{8\pi g_{0}^{2}}{v_{g}}\left[u_{C}(t)+D_{C,2}(t)\right]-\frac{2\pi g(t)g_{0}}{v_{g}}u_{A}(t)-\frac{4\pi g(t-2\tau)g_{0}}{v_{g}}D_{A,2}(t)-\frac{2\pi g(t-4\tau)g_{0}}{v_{g}}D_{A,4}(t)\nonumber\\
&&-\frac{6\pi g'(t-\tau)g_{0}}{v_{g}}D_{A,1}(t)-\frac{2\pi g'(t-3\tau)g_{0}}{v_{g}}D_{A,3}(t)-\frac{2\pi g_{0}^{2}}{v_{g}}\left[3D_{B,1}(t)+D_{B,3}(t)\right]. \label{eqc3}
\end{eqnarray}
which can be simplified to 
\begin{eqnarray}
\dot{u}_{A}(t)&=&-i\frac{4\pi g(t)g_{0}}{v_{g}}u_{B}(t)-i\frac{4\pi g'(t)g_{0}}{v_{g}}u_{C}(t), \label{eqc4}\\
\dot{u}_{B}(t)&=&-i\Delta u_{B}(t)-i\frac{4\pi g(t)g_{0}}{v_{g}}u_{A}(t)-i\frac{4\pi g_{0}^{2}}{v_{g}}u_{C}(t), \label{eqc5}\\
\dot{u}_{C}(t)&=&-i\Delta u_{C}(t)-i\frac{4\pi g'(t)g_{0}}{v_{g}}u_{A}(t)-i\frac{4\pi g_{0}^{2}}{v_{g}}u_{B}(t), \label{eqc6}
\end{eqnarray}
if $\phi=\pi/2$ and $\tau\rightarrow0$. By assuming cosine-type time-dependent couplings $g(t)=\Delta_{g}\cos{(\Omega t+\theta)}$ and $g'(t)=\Delta_{g}\cos{(\Omega t)}$ for atom $A$ with $\Omega\equiv\Delta$, one finally has 
\begin{eqnarray}
\dot{u}_{A}(t)&\simeq&-iG_{0}e^{i\theta}u_{B}(t)-iG_{0}u_{C}(t), \label{eqc7}\\
\dot{u}_{B}(t)&\simeq&-iG_{0}e^{-i\theta}u_{A}(t)-2i\Gamma_{0}u_{C}(t), \label{eqc8}\\
\dot{u}_{C}(t)&\simeq&-iG_{0}u_{A}(t)-2i\Gamma_{0}u_{B}(t), \label{eqc9}
\end{eqnarray}
which shows a protected all-to-all interaction with synthetic magnetic flux $\theta$. Having in mind that $G_{0}=\chi\Gamma_{0}$, directional excitation circulation can be expected if $\chi=2$ and $\text{mod}(\theta,\pi)=\pi/2$.

As mentioned in the main text, the protected all-to-all interaction among atoms $A$, $B$, and $C$ can also be implemented by using only one waveguide, provided that the coupling points of the three atoms are arranged according to the configuration in Fig.~\figpanel{fig1}{d}. In this case, we assume that the coupling points are equally spaced by $d'$ such that the phase accumulation (propagation time) of the field between adjacent coupling points becomes $\phi'=k_{0}d'$ ($\tau'=d'/v_{g}$). Again, atoms $B$ and $C$ are coupled to the waveguide with identical and constant strength $g_{0}$, while atom $A$ interacts with the waveguide with time-dependent strength $g(t)$ at each coupling point. After some algebra, the dynamical equations of the model can be obtained as
\begin{eqnarray}
 \dot{u}_{A}(t)&=&-\frac{2\pi g(t)}{v_{g}}\left\{2g(t)u_{A}(t)+2g(t-3\tau')D'_{A,3}(t)+g_{0}\left[2D'_{B,1}(t)+D'_{B,2}(t)+D'_{B,4}(t)\right.\right.\nonumber\\
 &&\left.\left.+D'_{C,1}(t)+2D'_{C,2}(t)+D'_{C,5}(t)\right]\right\},\label{eqc10}\\
\dot{u}_{B}(t)&=&-i\Delta u_{B}(t)-\frac{2\pi g_{0}}{v_{g}}\left\{2g_{0}u_{B}(t)+2g_{0}D'_{B,3}(t)+2g(t-\tau')D'_{A,1}(t)+g(t-2\tau')D'_{A,2}(t)\nonumber\right.\\
&&\left.+g(t-4\tau')D'_{A,4}(t)+g_{0}\left[2D'_{C,1}(t)+D'_{C,2}(t)+D'_{C,4}(t)\right]\right\}, \label{eqc11}\\
\dot{u}_{C}(t)&=&-i\Delta u_{C}(t)-\frac{2\pi g_{0}}{v_{g}}\left\{2g_{0}u_{C}(t)+2g_{0}D'_{C,3}(t)+g(t-\tau')D'_{A,1}(t)+2g(t-2\tau')D'_{A,2}(t)\nonumber\right.\\
&&\left.+g(t-5\tau')D'_{A,5}(t)+g_{0}\left[2D'_{B,1}(t)+D'_{B,2}(t)+D'_{B,4}(t)\right]\right\}, \label{eqc12}
 \end {eqnarray}
where $D'_{j,l}(t)=\text{exp}(il\phi')u_{j}(t-l\tau')\Theta(t-l\tau')$. When $\phi'=(2m+1/3)\pi$ and $\tau\rightarrow0$, the above three equations become
\begin{eqnarray}
\dot{u}_{A}(t)&=&-i\frac{4\pi g(t)g_{0}}{v_{g}}\sin{\left(\frac{\pi}{3}\right)}u_{B}(t)-i\frac{4\pi g(t)g_{0}}{v_{g}}\sin{\left(\frac{\pi}{3}\right)}u_{C}(t), \label{eqc13}\\
\dot{u}_{B}(t)&=&-i\Delta u_{B}(t)-i\frac{4\pi g(t)g_{0}}{v_{g}}\sin{\left(\frac{\pi}{3}\right)}u_{A}(t)-i\frac{4\pi g_{0}^{2}}{v_{g}}\sin{\left(\frac{\pi}{3}\right)}u_{C}(t), \label{eqc14}\\
\dot{u}_{C}(t)&=&-i\Delta u_{C}(t)-i\frac{4\pi g(t)g_{0}}{v_{g}}\sin{\left(\frac{\pi}{3}\right)}u_{A}(t)-i\frac{4\pi g_{0}^{2}}{v_{g}}\sin{\left(\frac{\pi}{3}\right)}u_{B}(t), \label{eqc15}
\end{eqnarray} 
which are identical with Eqs.~(\ref{eqc4})-(\ref{eqc6}), except for the modified effective coupling strengths. By assuming $g(t)=\Delta_{g}\cos(\Omega t+\theta)$ and performing the transformation $u_{B,C}(t)\rightarrow u_{B,C}(t)\text{exp}(-i\Delta t)$, Eqs.~(\ref{eqc13})-(\ref{eqc15}) become
\begin{eqnarray}
\dot{u}_{A}(t)&\simeq&-iG'e^{i\theta}u_{B}(t)-iG'e^{i\theta}u_{C}(t), \label{eqc16}\\
\dot{u}_{B}(t)&\simeq&-iG'e^{-i\theta}u_{A}(t)-2i\Gamma'u_{C}(t), \label{eqc17}\\
\dot{u}_{C}(t)&\simeq&-iG'e^{-i\theta}u_{A}(t)-2i\Gamma'u_{B}(t), \label{eqc18}
\end{eqnarray} 
where $G'=G_{0}\sin{(\pi/3)}=2\pi \Delta_{g}g_{0}\sin{(\pi/3)}/v_{g}$ and $\Gamma'=\Gamma_{0}\sin{(\pi/3)}=2\pi g_{0}^{2}\sin{(\pi/3)}/v_{g}$. Clearly, the effective coupling phase can always be gauged away via the transformation $u_{A}(t)\rightarrow u_{A}(t)\text{exp}(i\theta)$. Therefore, phase-dependent dynamics cannot be observed in this case. 
\end{widetext}

\section{Frequency-modulation scheme}\label{appd} 

In this Appendix, we consider that atoms $A$ and $B$ (recalling the giant-atom dimer) are coupled to the waveguide in the braided manner, yet with constant and uniform couplings (coupling strength $g_{0}$) instead. While the transition frequency $\omega_{0}$ of atom $A$ is assumed to be constant, we consider a frequency modulation for atom $B$ such that there is a small time-dependent detuning $\Delta_{0}+\Delta(t)$ between the two atoms. In this case, the Hamiltonian of the model can be written as
\begin{eqnarray}
H'&=&H_{\text{a}}'+H_{\text{w}}+H_{\text{int}}', \label{eqd1}\\
H_{\text{a}}'&=&\omega_{0}\sigma_{A}^{+}\sigma_{A}^{-}+[\omega_{0}+\Delta_{0}+\Delta(t)]\sigma_{B}^{+}\sigma_{B}^{-}, \label{eqd2}\\
H_{\text{int}}'&=&\int dk g_{0}\left[\left(1+e^{2ikd}\right)\sigma_{A}^{+}a_{k}\nonumber\right.\\
&&\left.+\left(e^{ikd}+e^{3ikd}\right)\sigma_{B}^{+}a_{k}+\text{H.c.}\right], \label{eqd3}
\end{eqnarray} 
where $H_{\text{w}}$ is identical with that in Eq.~(\ref{eq3}). With the single-excitation state of the system given in Eq.~(\ref{eq5}) and a similar calculation procedure as shown in Sec.~\ref{sec2}, one can obtain the dynamical equations of the atomic excitation amplitudes as
\begin{eqnarray}
\dot{u}_{A}(t)&=&-\Gamma_{0}\left[2u_{A}(t)+2D_{A,2}(t)+3D_{B,1}(t)\right.\nonumber\\
&&\left.+D_{B,3}(t)\right], \label{eqd4}\\
\dot{u}_{B}(t)&=&-i[\Delta_{0}+\Delta(t)]u_{B}(t)-\Gamma_{0}\left[2u_{B}(t)\nonumber\right.\\
&&\left.+2D_{B,2}(t)+3D_{A,1}(t)+D_{A,3}(t)\right], \label{eqd5}
\end{eqnarray}
where $\Gamma_{0}=2\pi g_{0}^{2}/v_{g}$ and $D_{j,l}(t)=\text{exp}(il\phi)u_{j}(t-l\tau)\Theta(t-l\tau)$ as defined in the main text. Once again, in the Markovian regime with negligible time delays and if $\phi=\pi/2$, the above two equations can be simplified to
\begin{eqnarray}
\dot{u}_{A}(t)&=&-2i\Gamma_{0}u_{B}(t), \label{eqd6}\\
\dot{u}_{B}(t)&=&-i[\Delta_{0}+\Delta(t)]u_{B}(t)-2i\Gamma_{0}u_{A}(t). \label{eqd7}
\end{eqnarray}
Now we consider a cosine-type modulation $\Delta(t)=\Delta_{g}'\cos{(\Omega' t+\theta')}$ (where $\Delta_{g}'$, $\Omega'$, and $\theta'$ are the amplitude, frequency, and initial phase of the modulation, respectively) for the detuning and perform a transformation 
\begin{equation}
u_{B}(t)\rightarrow u_{B}(t)e^{-i\Delta_{0}t}e^{-i\eta\sin{(\Omega't+\theta')}}
\label{eqd8}
\end{equation}
with $\eta=\Delta_{g}'/\Omega'$, Eqs.~(\ref{eqd6}) and (\ref{eqd7}) become
\begin{eqnarray}
\dot{u}_{A}(t)&=&-2i\Gamma_{0}u_{B}(t)e^{-i\Delta_{0}t}e^{-i\eta\sin{(\Omega't+\theta')}}, \label{eqd9}\\
\dot{u}_{B}(t)&=&-2i\Gamma_{0}u_{A}(t)e^{i\Delta_{0}t}e^{i\eta\sin{(\Omega't+\theta')}}. \label{eqd10}
\end{eqnarray}
Assuming $\Omega'=\Delta_{0}\gg2\Gamma_{0}$ and using the Jacobi-Anger expansion 
\begin{equation}
e^{-iz\sin{x}}=\sum_{q=-\infty}^{+\infty}J_{q}(z)e^{-iqx},
\label{eqd11}
\end{equation}
where $J_{q}(z)$ is the Bessel function of the first kind, one finally has
\begin{eqnarray}
\dot{u}_{A}(t)&\simeq&-2i\Gamma_{0}J_{-1}(\eta)u_{B}(t)e^{i\theta'}, \label{eqd12}\\
\dot{u}_{B}(t)&\simeq&-2i\Gamma_{0}J_{-1}(\eta)u_{A}(t)e^{-i\theta'} \label{eqd13}.
\end{eqnarray}
Clearly, a complex DFI between atoms $A$ and $B$ can also be created in this case.



\begin{thebibliography}{99}





























\bibitem {fiveyear} A.~F. Kockum, Quantum optics with giant atoms—the first five years, in \emph{International Symposium on Mathematics, Quantum Theory, and Cryptography} (Springer, Singapore, 2021), \href{https://link.springer.com/chapter/10.1007%2F978-981-15-5191-8_12}{pp: 125-146}.

\bibitem {LambAFK} A.~F. Kockum, P.~Delsing, and G.~Johansson, Designing frequency-dependent relaxation rates and Lamb shifts for a giant articial atom, \href{https://journals.aps.org/pra/abstract/10.1103/PhysRevA.90.013837}{Phys. Rev. A \textbf{90}, 013837 (2014)}.

\bibitem {GLZ2017} L.~Guo, A.~Grimsmo, A.~F. Kockum, M.~Pletyukhov, and G.~Johansson, Giant acoustic atom: A single quantum system with a deterministic time delay, \href{https://journals.aps.org/pra/abstract/10.1103/PhysRevA.95.053821}{Phys. Rev. A \textbf{95}, 053821 (2017)}.

\bibitem {oscillate} L.~Guo, A.~F. Kockum, F.~Marquardt, and G.~Johansson, Oscillating bound states for a giant atom, \href{https://journals.aps.org/prresearch/abstract/10.1103/PhysRevResearch.2.043014}{Phys. Rev. Res. \textbf{2}, 043014 (2020)}.

\bibitem {WXchiral1} X.~Wang, T.~Liu, A.~F. Kockum, H.-R. Li, and F.~Nori, Tunable Chiral Bound States with Giant Atoms, \href{https://journals.aps.org/prl/abstract/10.1103/PhysRevLett.126.043602}{Phys. Rev. Lett. \textbf{126}, 043602 (2020)}.

\bibitem {ZhaoWbound} W.~Zhao and Z.~Wang, Single-photon scattering and bound states in an atom-waveguide system with two or multiple coupling points, \href{https://journals.aps.org/pra/abstract/10.1103/PhysRevA.101.053855}{Phys. Rev. A \textbf{101}, 053855 (2020)}.

\bibitem {VegaPRA} C.~Vega, M.~Bello, D.~Porras, and A.~Gonz\'{a}lez-Tudela, Qubit-photon bound states in topological waveguides with long-range hoppings, \href{https://journals.aps.org/pra/abstract/10.1103/PhysRevA.104.053522}{Phys. Rev. A \textbf{104}, 053522 (2021)}.

\bibitem {YuanGA} H.~Xiao, L.~Wang, Z.-H.~Li, X.~Chen, and L.~Yuan, Bound state in a giant atom-modulated resonators system, \href{https://www.nature.com/articles/s41534-022-00591-7}{npj Quantum Info. \textbf{8}, 80 (2022)}.

\bibitem {TopoCheng} W.~Cheng, Z.~Wang, and Y.-x. Liu, Topology and retardation effect of a giant atom in a topological waveguide, \href{https://journals.aps.org/pra/abstract/10.1103/PhysRevA.106.033522}{Phys. Rev. A \textbf{106}, 033522 (2022)}.

\bibitem {2DtopoGA} C.~Vega, D.~Porras, and A.~Gonz\'{a}lez-Tudela, Topological multi-mode waveguide QED, \href{https://arxiv.org/abs/2207.02090}{arXiv:2207.02090}.

\bibitem {oscillate2} K.~H. Lim, W.-K. Mok, and L.-C. Kwek, Oscillating bound states in non-Markovian photonic lattices, \href{https://arxiv.org/abs/2208.11097}{arXiv:2208.11097}.

\bibitem {DLlambda} L.~Du and Y.~Li, Single-photon frequency conversion via a giant $\Lambda$-type atom, \href{https://journals.aps.org/pra/abstract/10.1103/PhysRevA.104.023712}{Phys. Rev. A \textbf{104}, 023712 (2021)}.

\bibitem {DLprr} L.~Du, Y.-T. Chen, and Y.~Li, Nonreciprocal frequency conversion with chiral $\Lambda$-type atoms, \href{https://journals.aps.org/prresearch/abstract/10.1103/PhysRevResearch.3.043226}{Phys. Rev. Res. \textbf{3}, 043226 (2021)}.

\bibitem {JiaGA1} Q.~Y. Cai and W.~Z. Jia, Coherent single-photon scattering spectra for a giant-atom waveguide-QED system beyond the dipole approximation, \href{https://journals.aps.org/pra/abstract/10.1103/PhysRevA.104.033710}{Phys. Rev. A \textbf{104}, 033710 (2021)}.

\bibitem {JiaGA2} S.~L. Feng and W.~Z. Jia, Manipulating single-photon transport in a waveguide-QED structure containing two giant atoms, \href{https://journals.aps.org/pra/abstract/10.1103/PhysRevA.104.063712}{Phys. Rev. A \textbf{104}, 063712 (2021)}.

\bibitem {YinScattering} X.-L. Yin, Y.-H. Liu, J.-F. Huang, and J.-Q. Liao, Single-photon scattering in a giant-molecule waveguide-QED system, \href{https://journals.aps.org/pra/abstract/10.1103/PhysRevA.106.013715}{Phys. Rev. A \textbf{106}, 013715 (2022)}.

\bibitem {ZhaoWScattering} W.~Zhao, Y.~Zhang, and Z.~Wang, Phase-modulated Autler-Townes splitting in a giant-atom system within waveguide QED, \href{https://journal.hep.com.cn/fop/EN/10.1007/s11467-021-1135-0}{Front. Phys. \textbf{17}, 42506 (2022)}.

\bibitem {CYTcp} Y.-T. Chen \emph{et al.}, Nonreciprocal and chiral single-photon scattering for giant atoms, \href{https://www.nature.com/articles/s42005-022-00991-3}{Commun. Phys. \textbf{5}, 215 (2022)}.

\bibitem {ZhuScattering} Y.~T. Zhu, S.~Xue, R.~B. Wu, W.~L. Li, Z.~H. Peng, and M.~Jiang, Spatial-nonlocality-induced non-Markovian electromagnetically induced transparency in a single giant atom, \href{https://journals.aps.org/pra/abstract/10.1103/PhysRevA.106.043710}{Phys. Rev. A \textbf{106}, 043710 (2022)}.

\bibitem {nonexp} G.~Andersson, B.~Suri, L.~Guo, T.~Aref, and P.~Delsing, Non-exponential decay of a giant artificial atom, \href{https://www.nature.com/articles/s41567-019-0605-6}{Nat. Phys. \textbf{15}, 1123 (2019)}.

\bibitem {LonghiGA} S.~Longhi, Photonic simulation of giant atom decay, \href{https://opg.optica.org/ol/abstract.cfm?uri=ol-45-11-3017&origin=search}{Opt. Lett. \textbf{45}, 3017 (2020)}.

\bibitem {DLretard} L.~Du, M.-R. Cai, J.-H. Wu, Z.~Wang, and Y.~Li, Single-photon nonreciprocal excitation transfer with non-Markovian retarded effects, \href{https://journals.aps.org/pra/abstract/10.1103/PhysRevA.103.053701}{Phys. Rev. A \textbf{103}, 053701 (2021)}.

\bibitem {LvGA} Q.-Y. Qiu, Y.~Wu, and X.-Y. L\"{u}, Collective radiance of giant atoms in non-Markovian regime, \href{https://link.springer.com/article/10.1007/s11433-022-1990-x}{Sci. China Phys., Mech. Astron. \textbf{66}, 224212 (2023)}.

\bibitem {AFKchiral} A.~Soro, and A.~F. Kockum, Chiral quantum optics with giant atoms, \href{https://journals.aps.org/pra/abstract/10.1103/PhysRevA.105.023712}{Phys. Rev. A \textbf{105}, 023712 (2022)}.

\bibitem {WXchiral2} X.~Wang and H.-R. Li, Chiral quantum network with giant atoms, \href{https://iopscience.iop.org/article/10.1088/2058-9565/ac6a04#qstac6a04app1}{Quantum Sci. Technol. \textbf{7}, 035007 (2022)}.

\bibitem {DLsyn} L.~Du, Y.~Zhang, J.-H. Wu, A.~F. Kockum, and Y.~Li, Giant Atoms in a Synthetic Frequency Dimension, \href{https://journals.aps.org/prl/abstract/10.1103/PhysRevLett.128.223602}{Phys. Rev. Lett. \textbf{128}, 223602 (2022)}.




\bibitem {NoriGA} A.~F. Kockum, G.~Johansson, and F.~Nori, Decoherence-Free Interaction between Giant Atoms in Waveguide Quantum Electrodynamics, \href{https://journals.aps.org/prl/abstract/10.1103/PhysRevLett.120.140404}{Phys. Rev. Lett. \textbf{120}, 140404 (2018)}.

\bibitem {braided} B.~Kannan \emph{et al.}, Waveguide quantum electrodynamics with superconducting artificial giant atoms, \href{https://www.nature.com/articles/s41586-020-2529-9}{Nature (London) \textbf{583}, 775-779 (2020)}.

\bibitem {FCdeco} A.~Carollo, D.~Cilluffo, and F.~Ciccarello, Mechanism of decoherence-free coupling between giant atoms, \href{https://journals.aps.org/prresearch/abstract/10.1103/PhysRevResearch.2.043184}{Phys. Rev. Res. \textbf{2}, 043184 (2020)}.

\bibitem {disDFI1} E.~Shahmoon and G.~Kurizki, Nonradiative interaction and entanglement between distant atoms, \href{https://journals.aps.org/pra/abstract/10.1103/PhysRevA.87.033831}{Phys. Rev. A \textbf{87}, 033831 (2013)}.

\bibitem {disDFI2} J.~S. Douglas, H.~Habibian, C.-L. Hung, A.~V. Gorshkov, H.~J. Kimble, and D.~E. Chang, Quantum many-body models with cold atoms coupled to photonic crystals, \href{https://www.nature.com/articles/nphoton.2015.57}{Nat. Photon. \textbf{9}, 326 (2015)}.

\bibitem {disDFI3} D.~E. Chang, J.~S. Douglas, A.~Gonz\'{a}lez-Tudela, C.-L. Hung, and H.~J. Kimble, Colloquium: Quantum matter built from nanoscopic lattices of atoms and photons, \href{https://journals.aps.org/rmp/abstract/10.1103/RevModPhys.90.031002}{Rev. Mod. Phys. \textbf{90}, 031002 (2018)}.

\bibitem {AB1959} Y.~Aharonov and D.~Bohm, Significance of electromagnetic potentials in the quantum theory, \href{https://journals.aps.org/pr/abstract/10.1103/PhysRev.115.485}{Phys. Rev. \textbf{115}, 485 (1959)}.

\bibitem {syn1} R.~O. Umucalılar and I. Carusotto, Artificial gauge field for photons in coupled cavity arrays, \href{https://journals.aps.org/pra/abstract/10.1103/PhysRevA.84.043804}{Phys. Rev. A \textbf{84}, 043804 (2011)}.

\bibitem {syn2} J.~Dalibard, F.~Gerbier, G.~Juzeli\={u}nas, and P.~\"{O}hberg, Colloquium: Artificial gauge potentials for neutral atoms, \href{https://journals.aps.org/rmp/abstract/10.1103/RevModPhys.83.1523}{Rev. Mod. Phys. \textbf{83}, 1523 (2011)}.

\bibitem {syn3} K.~Fang, Z.~Yu, and S.~Fan, Photonic Aharonov-Bohm Effect Based on Dynamic Modulation, \href{https://journals.aps.org/prl/abstract/10.1103/PhysRevLett.108.153901}{Phys. Rev. Lett. \textbf{108}, 153901 (2012)}.

\bibitem {syn4} K.~Fang, Z.~Yu, and S.~Fan, Realizing effective magnetic field for photons by controlling the phase of dynamic modulation, \href{http://dx.doi.org/10.1038/nphoton.2012.236}{Nat. Photon. \textbf{6}, 782 (2012)}.

\bibitem {syn5} K.~Fang, Z.~Yu, and S.~Fan, Experimental demonstration of a photonic Aharonov-Bohm effect at radio frequencies, \href{https://journals.aps.org/prb/abstract/10.1103/PhysRevB.87.060301}{Phys. Rev. B \textbf{87}, 060301 (2013)}.

\bibitem {syn6} N.~A. Estep, D.~L. Sounas, J.~Soric, and Andrea Al\`{u}, Magnetic-free non-reciprocity and isolation based on parametrically modulated coupled-resonator loops, \href{https://www.nature.com/articles/nphys3134}{Nat. Phys. \textbf{10}, 923 (2014)}.

\bibitem {syn7} V.~Peano, C.~Brendel, M.~Schmidt, and F.~Marquardt, Topological Phases of Sound and Light, \href{https://journals.aps.org/prx/abstract/10.1103/PhysRevX.5.031011}{Phys. Rev. X \textbf{5}, 031011 (2015)}.

\bibitem {syn8} M.~Schmidt, S.~Kessler, V.~Peano, O.~Painter, and F.~Marquardt, Optomechanical creation of magnetic fields for photons on a lattice, \href{https://opg.optica.org/optica/fulltext.cfm?uri=optica-2-7-635&id=321806}{Optica \textbf{2}, 635 (2015)}.

\bibitem {syn9} K.~Fang \emph{et al.}, Generalized non-reciprocity in an optomechanical circuit via synthetic magnetism and reservoir engineering, \href{https://www.nature.com/articles/nphys4009}{Nat. Phys. \textbf{13}, 465 (2017)}.

\bibitem {Roushan} P.~Roushan \emph{et al.}, Chiral ground-state currents of interacting photons in a synthetic magnetic field, \href{https://www.nature.com/articles/nphys3930}{Nat. Phys. \textbf{13}, 146 (2017)}.

\bibitem {JinPRL} L.~Jin and Z.~Song, Incident Direction Independent Wave Propagation and Unidirectional Lasing, \href{https://journals.aps.org/prl/abstract/10.1103/PhysRevLett.121.073901}{Phys. Rev. Lett. \textbf{121}, 073901 (2018)}.

\bibitem {JiaST} X.~Guan, Y.~Feng, Z.-Y. Xue, G.~Chen, and S.~Jia, Synthetic gauge field and chiral physics on two-leg superconducting circuits, \href{https://journals.aps.org/pra/abstract/10.1103/PhysRevA.102.032610}{Phys. Rev. A \textbf{102}, 032610 (2020)}.

















\bibitem {moduscheme} Y.~Yin \emph{et al.}, Catch and Release of Microwave Photon States, \href{https://journals.aps.org/prl/abstract/10.1103/PhysRevLett.110.107001}{Phys. Rev. Lett. \textbf{110}, 107001 (2013)}.

\bibitem {footnote} In this paper, we always assume $2\pi g_{0}^{2}\tau/v_{g}\ll1$ so that the time delay $\tau$ is much smaller than the relaxation time of the giant atoms and the durations of the dynamical evolutions of our concern.

\bibitem {arxivClerk} A.~A. Clerk, Introduction to quantum non-reciprocal interactions: from non-Hermitian Hamiltonians to quantum master equations and quantum feedforward schemes, \href{https://scipost.org/10.21468/SciPostPhysLectNotes.44}{SciPost Phys. Lect. Notes, 44 (2022)}.

\bibitem {WXNJP} X.~Wang, H.-R. Li, and F.-L. Li, Generating synthetic magnetism via Floquet engineering auxiliary qubits in phonon-cavity-based lattice, \href{https://iopscience.iop.org/article/10.1088/1367-2630/ab776e}{New J. Phys. \textbf{22}, 033037 (2020)}.

\bibitem {James2007} D.~F. James and J.~Jerke, Effective Hamiltonian theory and its applications in quantum information, \href{https://cdnsciencepub.com/doi/10.1139/p07-060}{Can. J. Phys. \textbf{85}, 625 (2007)}.

\bibitem {GAcollision} D.~Cilluffo, A.~Carollo, S.~Lorenzo, J.~A. Gross, G.~M. Palma, and F.~Ciccarello, Collisional picture of quantum optics with giant emitters, \href{https://journals.aps.org/prresearch/abstract/10.1103/PhysRevResearch.2.043070}{Phys. Rev. Res. \textbf{2}, 043070 (2007)}.

\bibitem {DLprr2} L.~Du, Y.-T. Chen, Y.~Zhang, and Y.~Li, Giant atoms with time-dependent couplings, \href{https://journals.aps.org/prresearch/abstract/10.1103/PhysRevResearch.4.023198}{Phys. Rev. Res. \textbf{4}, 023198 (2022)}.





\bibitem {JTShen2005} J.-T. Shen and S.~Fan, Coherent Single Photon Transport in a One-Dimensional Waveguide Coupled with Superconducting Quantum Bits, \href{https://journals.aps.org/prl/abstract/10.1103/PhysRevLett.95.213001}{Phys. Rev. Lett. \textbf{95}, 213001 (2005)}. 

\bibitem {JTShen2009} J.-T. Shen and S.~Fan, Theory of single-photon transport in a single-mode waveguide. I. Coupling to a cavity containing a two-level atom, \href{https://journals.aps.org/pra/abstract/10.1103/PhysRevA.79.023837}{Phys. Rev. A \textbf{79}, 023837 (2009)}. 

\end{thebibliography}
\end{document}